# Open-path Dual-comb Spectroscopy for Multispecies Trace Gas Detection in the 4.5 µm to 5 µm Spectral Region

*Fabrizio R. Giorgetta[1,2], Jeff Peischl[3,4], Daniel I. Herman[1,2], Gabriel Ycas[1,2], Ian Coddington[5], Nathan R. Newbury[5], Kevin C. Cossel[5]\**

[1]Associate of the National Institute of Standards and Technology, 325 Broadway, Boulder, CO 80305, United States of America

[2]Department of Physics, University of Colorado, Boulder, CO 80309, United States of America

[3]Cooperative Institute for Research in Environmental Sciences, University of Colorado Boulder, Boulder, CO 80309, United States of America

[4]National Oceanic and Atmospheric Administration, Chemical Sciences Laboratory, Boulder, CO 80305, United States of America

[5]National Institute of Standards and Technology, Applied Physics Division, 325 Broadway, Boulder, CO 80305, United States of America

\* kevin.cossel@nist.gov

Open-path dual-comb spectroscopy provides multi-species atmospheric gas concentration measurements with high precision. Here, we extend open-path dual comb spectroscopy to the mid-infrared 5-µm atmospheric window, enabling atmospheric concentration retrievals of the primary greenhouse gases, $N_2O$, $CO_2$, and $H_2O$ as well as the criterion air pollutants $O_3$ and CO across 600 m and 2 km round-trip paths. We demonstrate measurements over a five-day period at two-minute temporal resolution with 80% uptime. The achieved precision is sufficient to resolve the atmospheric concentration variations of the multiple gas species; retrieved dry mixing ratios of CO and $N_2O$ are in good agreement with a co-located point sensor. In addition, the retrieved ratio of excess CO vs $CO_2$ agrees with similar urban studies but disagrees with the US National Emission Inventory by a factor of 3. Our retrieved ratio of excess $N_2O$ vs $CO_2$ exhibits a plume-dependent value, indicating the variability of sources of the greenhouse gas $N_2O$.

## 1. Introduction

The 4.5 µm to 5 µm spectral region can be used to measure a number of important atmospheric trace gases including the greenhouse gases $CO_2$ and $N_2O$ and the criteria air pollutants $O_3$ and CO. In addition, this spectral region is relatively clear of strong water absorption features, which allows for measurements across long open paths. Existing sensing technologies in this spectral region typically only measure one or two species; however, multispecies detection can be very beneficial for attributing sources. Open-path Fourier-transform infrared spectroscopy (OP-FTIR) can measure multiple gas species [1], but the low spectral resolution (typically > ~6 GHz) leads to potential issues with biases [2,3] and very long open paths are challenging due to the incoherent light source's divergence. Dual-comb spectroscopy (DCS) [4] is an attractive platform for open-path measurements of multiple atmospheric species [5] that can overcome these limitations. Like FTIR, the broad spectral coverage of DCS allows for simultaneous quantification of many trace gas species as well as the path-averaged temperature. However, DCS has higher spectral resolution and negligible instrument lineshape compared to OP-FTIR instruments enabling trace gas concentration measurements at higher precision [6]. In addition, the comb lasers emit a bright, single transversal mode beam which can be propagated long distances, day or night, enabling continuous observation of gas concentrations and fluxes with high precision and over large areas [7]. These hectometer to kilometer-scale open path lengths reduce the sensitivity of the measurements to wind field errors and thereby reduce errors when quantifying emission rates of upwind sources. Lastly, these path lengths are closely matched to the grid size of high-resolution (i.e., meso- and microscale) numerical weather prediction and air quality models, which reduces the model-measurement representation error [8].

In previous work, an open-path DCS system in the near-infrared at 1.6 µm was used to simultaneously measure $CH_4$, $CO_2$, $H_2O$ and air temperature [5,6] and the unique strengths of open-path DCS have enabled new measurement approaches for quantification of city scale

CO$_2$ emissions [7], detection of leaks in oil and gas infrastructure [9,10], emissions from cattle feed lots [in review], and vertical/horizontal trace-gas profiling [11]. Here, our interest is to move into the 4.5 µm to 5 µm atmospheric window, which can greatly expand the multi-species measurement capabilities of DCS. Previous lab-based DCS systems that can potentially access this spectral band have been based on optical parametric oscillators [12–15], interband and quantum cascade lasers [16,17], and difference frequency generation (DFG) using supercontinua [18–22]; however, only a few open-path DCS measurements have been demonstrated in the mid-infrared and all have been around 3 µm [15,19,20,23]. Here, we show a DFG-based, open-path dual-comb spectrometer capable of covering the 4.5 µm to 5 µm spectral region with comb-tooth resolution and sufficient output power for operation over long open paths. We demonstrate simultaneous measurements of H$_2$O, N$_2$O, CO$_2$, CO, and O$_3$ across 600 m and 2 km long paths (one-way distances of 300 m and 1 km). By measuring three critical atmospheric species not accessible to near-infrared DCS systems – CO, O$_3$ and N$_2$O – this spectrometer will enable future applications in understanding and monitoring urban air quality and greenhouse gas (GHG) emissions.

CO is important for air quality because it is a toxic gas and because it contributes to catalytic ozone production and destruction [24]. Globally, it also is the primary sink for the OH radical, which is the primary oxidant in the atmosphere and impacts the concentrations and distribution of greenhouse gases and pollutants. In addition, since CO is produced by incomplete combustion, it serves as a good tracer of anthropogenic emissions. For example, the ratio of CO to CO$_2$ can be used to track CO$_2$ emissions from fossil-fuel combustion and to distinguish different sources of combustion based on the combustion efficiency (e.g., power plants have a low ratio of CO to CO$_2$ compared to much less efficient nonroad vehicles) [25–27]. In urban areas, the ratio of CO/CO$_2$ has been used to track diurnal variations in CO$_2$ source contributions [25] and to perform some CO$_2$ source sector attribution [26]. Furthermore, measurements of both CO and CO$_2$ in fire plumes can determine the modified combustion efficiency, which is a measure of the total carbon consumed and is important for better understanding the trace gas emissions from the fire [28,29]. Finally, because of the relatively long tropospheric lifetime (days to weeks) [24], CO also provides a tracer of long-range atmospheric transport, for example, from wildfires and urban areas [27,30]. Thus, open-path monitoring of CO, combined with numerical weather models, could be used help identify the influence of wildfires, anthropogenic sources, and biogenic sources on urban air quality.

Ground-level $O_3$ is a major health hazard [31] and also has adverse impacts on vegetation and agriculture. In many regions, ground-level $O_3$ concentrations frequently exceed government control levels, thus there is a major emphasis on controlling and reducing $O_3$ levels. Tropospheric $O_3$ primarily arises from complex photochemical reactions of precursor species such as nitrogen oxides (NO and $NO_2$) as well as volatile organic compounds. The complex chemistry leads to significant challenges in developing a full understanding of ozone formation, especially in urban areas. Because of this complexity, state-of-the-art models still show persistent biases and often struggle to replicate high-$O_3$ events [32]. This can result in errors in exposure assessments and also complicate efforts to mitigate $O_3$. In many areas, wildfires and biomass burning are a potentially significant $O_3$ source; however, there is still uncertainty about the overall $O_3$ contribution from fires [30]. Simultaneous measurements of $O_3$ and CO could help to decrease this uncertainty. Furthermore, $O_3$ concentrations can have spatial gradients on the kilometer [33,34] and even tens-of-meter scale [35]. These gradients can lead to discrepancies between point measurements and models and their impact could be minimized with the long paths possible with DCS.

$N_2O$ is the third most prevalent anthropogenic GHG. Measurement of $N_2O$ is particularly challenging because the typical enhancements relative to atmospheric background levels are small (~1% or less). Thus, compared to $CO_2$ and $CH_4$, the sources of $N_2O$ are significantly less well known as are the magnitudes of the feedback cycles influencing the sources [36]. The primary natural and anthropogenic sources arise from microbial activity in soils and can be driven by crop overfertilization [36,37]. Quantifying these emissions is challenging due to their large temporal and spatial variability [38], but reliable monitoring of these emissions could improve agricultural efficiency while reducing environmental impact [37]. In urban areas, $N_2O$ is primarily emitted from vehicles; however, $N_2O$ is not routinely monitored in urban areas, so there are very few checks of emissions inventories. As with CO, the ratio of $N_2O$ to other gases can assist with source attribution. Improved determination of sector-specific emission ratios would allow high-resolution $CO_2$ emission inventories to be extended to other GHGs and criteria pollutants [39].

Below, we first describe the 4.5 µm to 5 µm open-path DCS system, which probed both a 2 km and 600 m roundtrip open path. In particular, we operated the system over the 600-m path for 5 days to compare the results with several *in situ* point sensors. We present precision analysis (Allan deviation) for the four species, $N_2O$, $CO_2$, CO, and $O_3$, and discuss potential systematic bias. Finally, we present example results analyzing the correlations of CO and $N_2O$ with $CO_2$ for several plumes detected during the five days of measurement. While the system here is configured to target CO and $N_2O$, the optical backbone of the system follows that of a near infrared DCS, which can precisely measure $CO_2$ and $CH_4$. Therefore, with some additional optical reconfiguration, a dual-channel near/mid-IR could potentially monitor all urban GHG emissions.

## 2. Experimental Setup

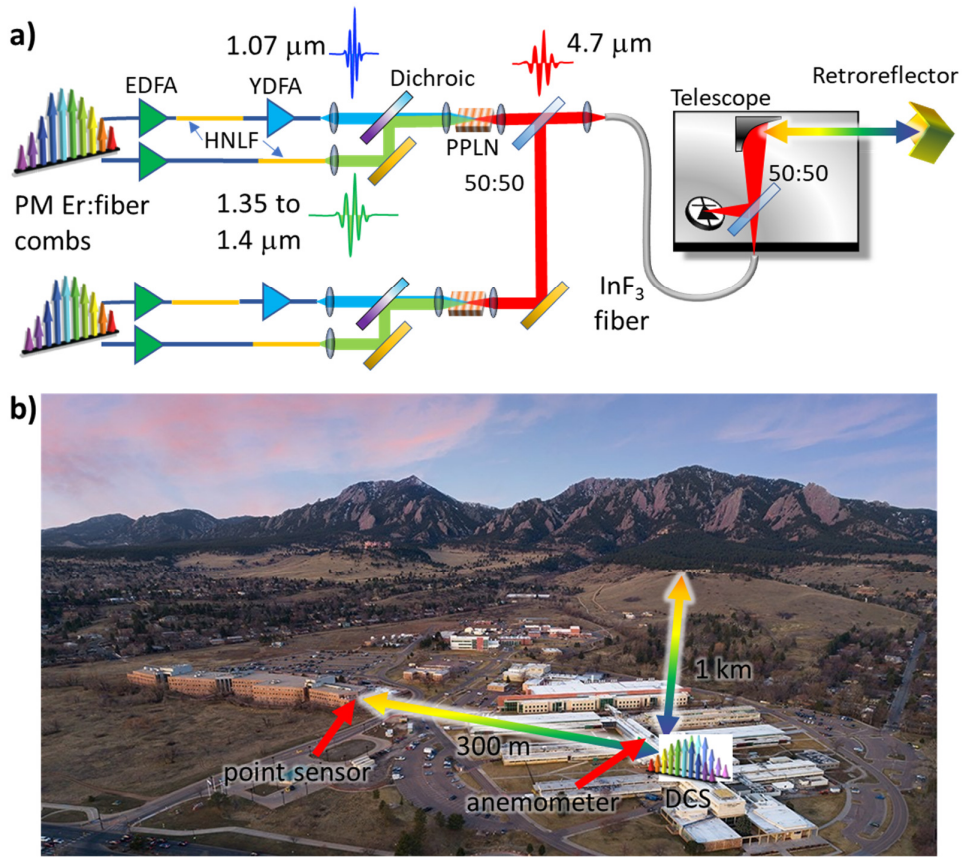

**Figure 1.** Overview. (a) Schematic of open-path DCS system and telescope. PM: Polarization-maintaining. EDFA: Erbium-doped fiber amplifier. HNLF: highly non-linear fiber. YDFA: Ytterbium-doped fiber amplifier. PPLN: periodically poled lithium niobate. (b) The system is located in a room atop the NIST building in Boulder, Colorado, and the light is sent to a retroreflector located on a balcony of the NOAA building ~300 m away. Alternatively, a second beam path launches the comb light to a reflector ~1 km away. A point sensor at the NOAA building also records $N_2O$, CO and $H_2O$ concentrations. Wind speed and direction are recorded with a 3D sonic anemometer located on the roof of the NIST building.

Figure 1 shows an overview of the measurement setup. The dual-comb spectrometer follows the basic design discussed in [19,20]. It is based on two fully stabilized Er:fiber frequency combs with repetition rates of ~200 MHz that are offset by 104 Hz. About 5 mW to10 mW of mid-infrared light covering 4500 nm to 4900 nm is generated by non-linear difference frequency generation between light at 1.07 µm and light at 1.35 µm to 1.4 µm originating from one frequency comb as shown in Figure 1. The 1.07 µm light is generated by amplification of spectrally broadened light using a Yb-doped fiber amplifier. The 1.35 µm to 1.4 µm light is obtained by spectral broadening using highly nonlinear fiber. The mid-infrared light from both combs is combined on a free-space beamsplitter and coupled into a single-mode $InF_3$ fiber, which runs to a telescope (Fig. 1a). At the telescope, light is launched from the fiber tip, passes through a 50:50 beamsplitter and is collimated by an off-axis parabolic mirror (180 mm focal length, 10 cm diameter). This collimated light is sent through the atmosphere to a 12.5 cm diameter retroreflector located 300 m or 1 km away. After reflection, the light traverses the same path, reflects off the telescope's 50:50 beamsplitter, and is collected on a 250 MHz mercury cadmium telluride (MCT) detector.

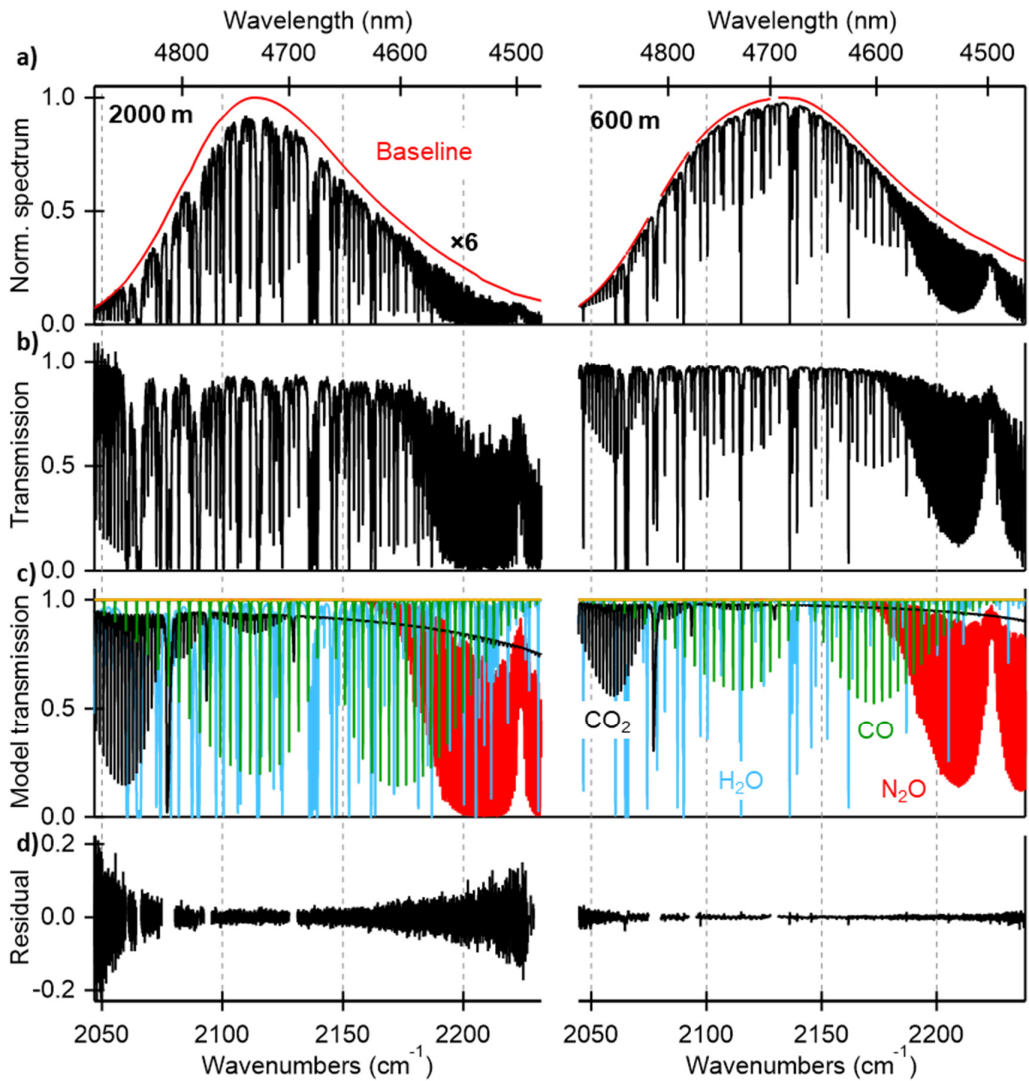

**Figure 2.** a) Measured return spectra for the 2000-m and 600-m round-trip paths (black line). The overall shape reflects the comb spectra and is modeled as the baseline, or zero-absorption spectrum, shown as the red line. The strong atmospheric absorption appears as many deep narrow lines. b) Transmission, obtained by division of the measured spectrum with the baseline. c) Model transmission for $CO_2$, $H_2O$, CO and $N_2O$ for concentrations obtained by fitting the model to the transmission in b); The $O_3$ transmission is below 1 % for both paths.
d) Residual obtained by subtracting the model (c) from the measured transmission (b).

For comparison between DCS and a calibrated point sensor, an inlet was located at the retroreflector end of the 300 m path, as shown in Fig. 1b. This inlet was connected to an off-axis cavity-enhanced spectrometer run by the National Oceanic and Atmospheric Administration (NOAA). This instrument measures $N_2O$, CO, and $H_2O$ with a 1 Hz measurement rate (averaged to 1 minute). Two standard gases were regularly delivered to the instrument inlet line throughout the five-day measurement period to evaluate instrument sensitivity between 317 and 385 ppb (nmol/mol) $N_2O$ and 58 and 990 ppb CO. The standards were calibrated after the study using $N_2O$ and CO standard tanks tied to the WMO standards WMO-N2O-X2006A and WMO-CO-X2014 [40]. We estimate a total uncertainty of ±0.5 ppb for $N_2O$ and ±0.8 ppb for CO. For $O_3$ there was no co-located point sensor. Instead, below we compare with data from a Colorado Department of Health and Environment (CDPHE) sensor located about 15 km north of the DCS open path that provided $O_3$ at 1-hour temporal resolution. During this measurement campaign, there were no nearby operational $CO_2$ point sensors. Finally, the wind direction and speed were measured at 10 Hz with a 3D sonic anemometer located at the NIST building as shown in Fig. 1b.

The DCS system samples the atmosphere and generates a time-domain interferogram every 9.6 ms, as determined by the 104 Hz offset in the comb repetition rates. The interferograms are phase corrected in real time and then co-added on a field-programmable gate array (FPGA) to give an averaged spectrum every ~2 minutes [19,23], as shown in Fig. 2. The resulting spectra are then analyzed to retrieve the path-averaged concentrations [5,6]. To do this, we must first divide the measured spectrum by the zero-absorption dual-comb spectrum, referred to as the baseline spectrum. Because the gas absorption spectra consist of narrow lines, we determine the slowly varying baseline by fitting the measured spectrum to a piecewise polynomial function that also includes the gas absorption profiles [5–7]. (The $CO_2$ concentration is held fixed in this fit at its nominal value since its spectrum in this region includes a broad continuum component.) The measured spectrum is then divided by this baseline to yield the open-path transmission spectrum. The transmission spectrum is fit with a gas absorption model to determine the path-averaged concentration of each gas. The spectral parameters for this model were obtained from HITRAN 2016 [41]. For gases other than water, the fitted concentration is then corrected to the dry concentration without water present.

## 3. Results

Figure 2 shows the measured path-averaged concentrations of $N_2O$, $CO$, $CO_2$, $H_2O$, and $O_3$ over 5 days of near-continuous measurements across the 600 m roundtrip path at 2-minute time resolution. In addition to the trace gas species, the DCS instrument retrieves the path-averaged temperature based on the fit. Also shown for comparison are the data from the point sensor located at NOAA for CO, $N_2O$, and $H_2O$ (in blue) and from the CDPHE point sensor for $O_3$ (in red). As discussed below, the DCS $O_3$ data is offset by +10 ppb in Figure 2. For most of the gases, we observe significant fast and slow variability in concentrations due to the combination of atmospheric effects such as changing wind and planetary boundary layer height, and plumes from sources intersecting the beam path. During those time periods, $O_3$ varies between 0 ppb to 50 ppb with noticeable diurnal variation arising from the fact that $O_3$ production is dominated by photochemical processes. The other criteria pollutant measured, CO, reaches a background level of ~100 ppb but shows large spikes up to ~8× background due to local sources. Unlike the criteria pollutants, the GHGs $CO_2$ and $N_2O$ show smaller variations relative to background highlighting the need for high precision. This is especially true for $N_2O$ where the background level is ~330 ppb and the enhancements are only about 1% to 2% of the background. For $CO_2$, the background level reached ~420 ppm (µmol/mol) with the largest spike ~25% of the background.

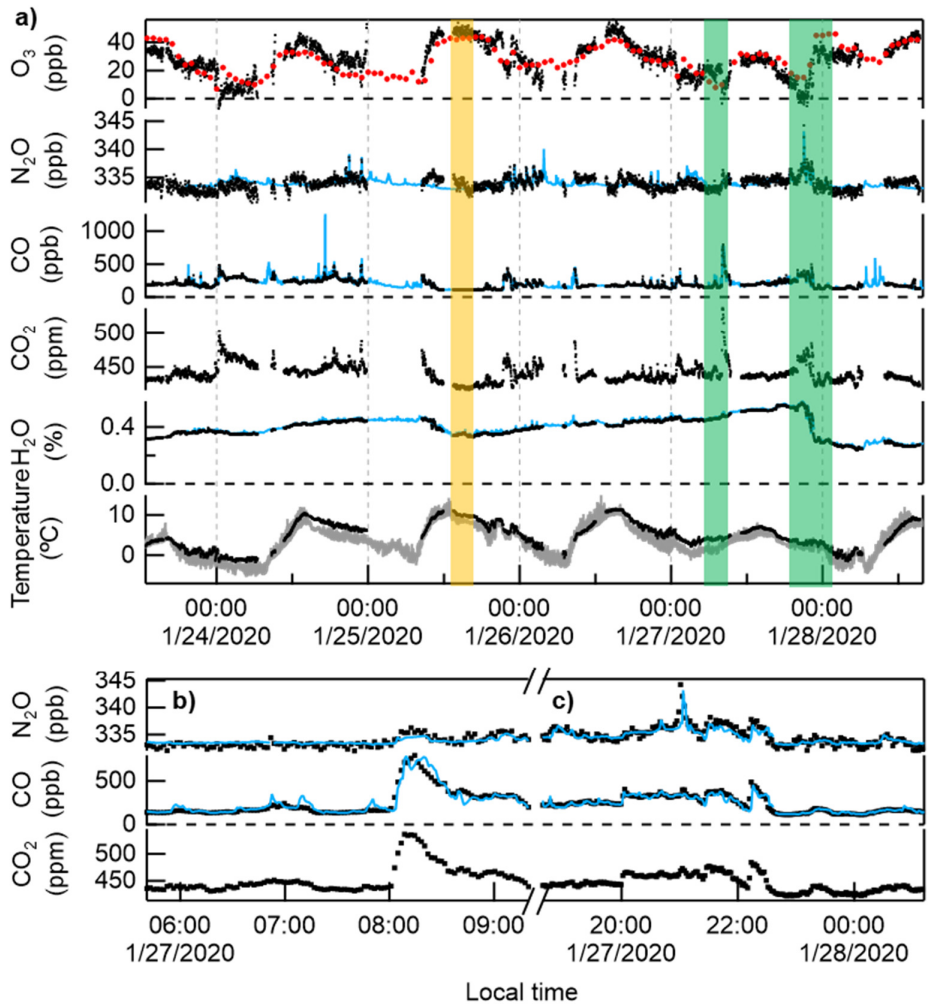

**Figure 3.** (a) Retrieved path-averaged concentration and temperature for the DCS measurements (black) compared to measurements from in situ sensors (NOAA, blue; CDPHE, red dots; anemometer temperature, grey). Note that the retrieved $O_3$ concentration has been offset by 10 ppb. The orange-highlighted time period is used for the Allan deviation given in Fig. 4. (b-c) An expanded view of the green highlighted regions in part (a), which are used for the $N_2O$ vs $CO_2$ correlation plots in Figure 5.

We first use these data to evaluate the performance of the open-path DCS instrument for both sensitivity and bias. In order to determine the detection sensitivity for a trace gas, we evaluate the Allan-Werle deviation [42] over a time period with relatively constant concentrations when atmospheric variability is lower because the system is primarily measuring relatively clean air from the west. (See orange highlighting in Fig. 3.) As seen in Figure 4, the Allan deviations average down linearly with the square-root of averaging time for short averaging times until reaching a floor, with the exception of CO. For CO, we attribute the roughly constant 0.4 ppb sensitivity to true atmospheric variability, as it is consistent with the Allan deviation from the NOAA point sensor over the same time period. For $CO_2$, we also attribute the floor of the Allan deviation at 0.6 ppm to atmospheric variability, as it is consistent with similar data for $CO_2$ acquired with a near-infrared open-path DCS [6] at the same location. For $O_3$, we again attribute the 0.6 ppb floor to atmospheric variability as it is similar to the point-to-point variation measured with by the CDPHE sensor at 1-hour resolution. For $N_2O$, the Allan deviation reaches about 0.3 ppb and then appears to increase slightly at longer averaging times. In contrast, the point sensor at NOAA shows an Allan deviation that is flat at around 0.01 ppb. We attribute the ~0.3 ppb limit to the DCS sensitivity for $N_2O$ primarily to a small (<0.5%) time varying bias from nonlinearities discussed below. After correcting for this bias, the Allan deviation (gray) shows the expected linear trend.

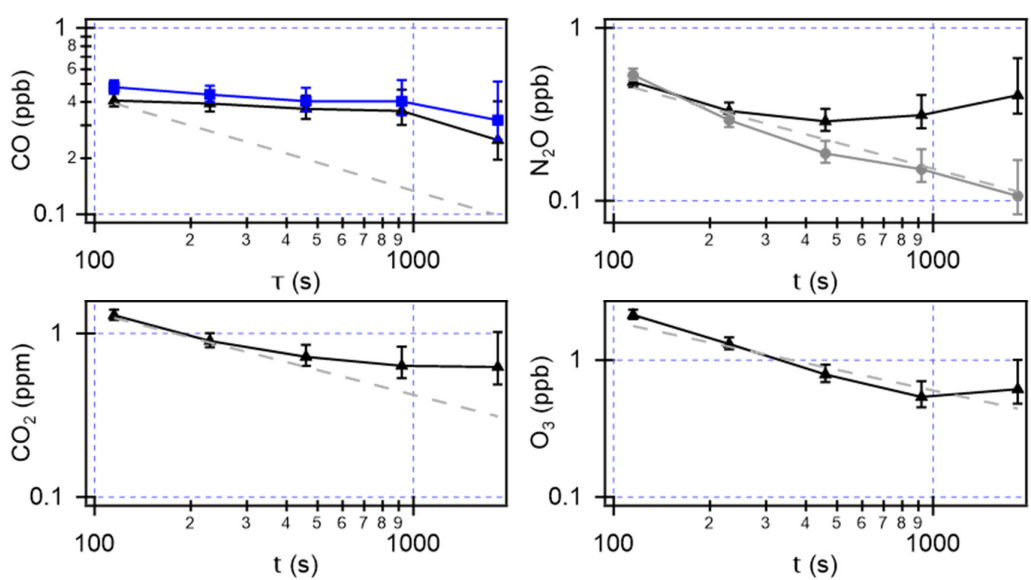

**Figure 4.** Allan-Werle deviation plots for the DCS measurements for CO, $N_2O$, $CO_2$, and $O_3$. Grey-dashed lines indicate the expected slope for white noise. The additional blue trace for CO is for the *in situ* point sensor. The additional grey trace for $N_2O$ is the DCS data after correction for variation due to detector nonlinearity.

Table 1 summarizes primary fitted gas species, their peak absorption, typical measured concentration, and the DCS sensitivity for both the 600-m and 2-km paths. Each molecule's spectral signature is shown in Fig. 2. For the 2-km path, we did not observe over a long enough time period with low atmospheric variability for an Allan deviation, so the sensitivity is the standard deviation of a short, detrended time series. In the ideal case, the sensitivity would improve linearly with path length; however, the detection SNR is about four times lower for the 2-km path due to reduced return power, so the sensitivity for the 2-km path is slightly worse than that of the 600-m path. In addition to the main species listed in Table 1, we are also able to detect the isotopologues $^{13}CO$, $^{13}CO_2$, $^{14}N^{15}NO$, $H_2^{18}O$, and $H_2^{17}O$ with δ-value precisions of 50‰, 5‰, 30‰, 20‰, and 60‰, respectively. Ratios of stable isotopes provide valuable information about the gas sources and sinks [43]. Open-path measurements of isotope ratios could be especially beneficial for water isotopologues because of the challenges associated with closed path sampling systems; however, currently the isotope ratio sensitivity of the open-path DCS is not high enough to be very useful for atmospheric measurements (typically on the order of 1‰ is desired). Improvements to the return power and detector noise as well as longer averaging time should enable future isotope ratio measurements at the 1‰ level.

Table 1. Summary of fitted species and sensitivity for both measurement paths.
*Concentration given in mole fraction: ppm = μmol/mol, ppb = nmol/mol. ^: Sensitivity likely limited by atmospheric variability and not instrument noise.

| Gas | Spectral region (cm$^{-1}$) | Peak absorption (600m) | Average concentration* | Sensitivity @ 600m and 2 minutes | Sensitivity @ 2km and 2 minutes |
|---|---|---|---|---|---|
| $N_2O$ | 2150-2250 | 80% | 334 ppb | 0.5 ppb | 0.88 ppb |
| CO | 2075-2225 | 50% | 202 ppb | 0.4 ppb^ | 0.6 ppb^ |
| $CO_2$ | 2040-2100 | 40% | 440 ppm | 1.5 ppm | 1.9 ppm |
| $O_3$ | 2075-2130 | 0.4% | 28 ppb | 2 ppb | 2.3 ppb |
| $H_2O$ | 2040-2250 | 100% | 0.4% | 11 ppm^ | 56 ppm^ |
| Temp | full | n/a | n/a | 0.1 C | 0.1 C |

We also investigated potential sources of bias in the DCS measurements. To do this, we first compare the DCS measurements with the NOAA point sensor. As evident from the time series in Fig 2, the point sensor and DCS are well correlated. Their difference yields small, constant offsets (DCS – point sensor) of 1.8 ppb, -0.88 ppm, and -2.9 ppb for CO, $H_2O$, and $N_2O$, respectively, corresponding to percentage offsets of 0.9%, -0.02%, and -0.9%. These offsets are all within the known uncertainties of the HITRAN line parameters. We also evaluated the correlation between the DCS and NOAA sensors for CO, $H_2O$, and $N_2O$. The slope was 0.985 for CO, 0.991 for $H_2O$, and 1.0 for $N_2O$ and with Pearson correlation coefficients of 0.91, 0.996, and 0.5, respectively. Again these slopes are well within the uncertainties of the HITRAN database.

We can conduct a similar analysis for $O_3$ between the DCS measurements and the CDPHE point sensor. However, the different measurement timescales (2-minute vs 1-hour) and the 15-km separation in locations make a direct comparison difficult. As noted earlier, we do find an approximate -10-ppb offset to the $O_3$ measurements from the DCS, which was corrected in Figure 2. This offset likely occurs because the $O_3$ absorption signal is very weak and fairly broad, which makes it susceptible to interference from inaccurate lineshape models for the stronger species. Improved lineshape models or the inclusion of a persistent residual structure in the fit would help to reduce the $O_3$ offset.

For the $N_2O$ retrievals, a small, time-varying bias does seem to be present in the data. This can be seen, for example, in the orange highlighted region in Fig. 2 where the DCS measurements shows more slow variability than the point sensor. These temporal variations are strongly correlated with the received optical power and are attributed to non-linearities in the photodetector. By correcting for the temporal power variations, we were able to significantly reduce the $N_2O$ variation during the time period used for the Allan deviation analysis (compare the gray and black data in Fig. 3b). This explanation is further supported by laboratory measurements of $N_2O$ in a cell, where a similar level bias could be reproduced by changing the detected optical power. This bias is low; we find a 0.6% change in $N_2O$ with a factor of two change in received power. Nevertheless, this bias is significant for $N_2O$ because of the low atmospheric variability. While this effect is possibly present for other gases as well, it is not significant compared to the atmospheric variability. In the future, this bias could be calibrated through laboratory measurements, and then a correction applied based on the measured received power, or the effect could be removed by using a lower nonlinearity MCT receiver.

## 4. Atmospheric Science Discussion

As discussed earlier, one of the significant advantages of multi-species detection is the ability to measure correlations between species. First, we focus on CO and $CO_2$. From Figure 2, it is already apparent that these two species are strongly correlated with both exhibiting a slowly varying background punctuated by larger spikes. This strong correlation is expected as they are both emitted from combustion sources, and the measurement site is located close to major roads. The variable background arises from atmospheric effects such as changing planetary boundary layer height, which effectively traps the emitted CO and $CO_2$ in a shallower or deeper box. The larger spikes are a result of gas plumes passing through the open beam path. To analyze these plumes, we calculate the excess concentrations, $CO_{xs}$ and $CO_{2xs}$, above a background level, as calculated using the robust baseline estimator approach [44]. As shown in Figure 4, $CO_{xs}$ and $CO_{2xs}$ are correlated with a slope of 5.28(4) ppb CO/ppm $CO_2$ ($R^2$ of 0.86). The tight correlation indicates that the combustion sources from these plumes have similar combustion efficiencies.  Interestingly, we did observe a single plume event with a significantly higher $CO_{xs}/CO_{2xs}$ ratio. Based on the wind and plume behavior, we believe that it is a local source, but the specific source is uncertain.

The expected ratio of CO vs $CO_2$ depends on the efficiency of combustion for a given source. Recent studies in urban areas have observed values between 4 ppb/ppm to 8 ppb/ppm [26,39,45,46], consistent with our value of 5.28 ppb/ppm. However, we do not find similar consistency between our measured ratio and one calculated from the US EPA National Emission Inventory (NEI 2017). If we just consider on-road mobile sources in Boulder County, the NEI yields a $CO/CO_2$ ratio of 17.9 ppb/ppm, which is significantly higher than the observed ratio (including additional sources beyond on-road mobile sources only further increases the ratio). This discrepancy is consistent with previous studies [26,39,45,46], which have suggested decreasing the NEI CO inventory by a factor of 2 to 3 to improve agreement with observations, although none of these previous studies were also in the same Denver area. Here, we find that the NEI overestimates CO by a slightly higher amount (about 3.4×), but without a longer duration data acquisition it is difficult to estimate an uncertainty on this scaling. Furthermore, we do not have enough data to attribute whether any discrepancy is due to incorrect CO or $CO_2$ emissions in the NEI.

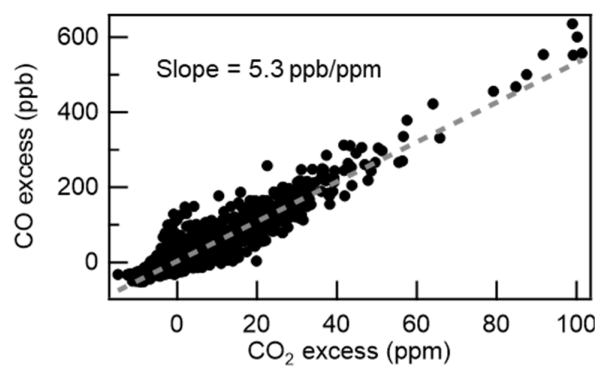

**Figure 5.** Correlation between excess CO and excess $CO_2$. The line fit is shown by the grey dashed line.

Next, we consider the $N_2O$ sources. We expect that the observed $N_2O$ enhancements arise primarily from vehicles where it is produced in catalytic converters, since any agricultural contribution is expected to be small in January. In the case of vehicle emissions, $N_2O$ should be correlated with $CO_2$, just as with CO. However, traffic sources have a wide range of $N_2O/CO_2$ depending on the type of fuel source, type of catalytic converter, age of catalytic converter, and driving condition with values range from 0.004 ppb/ppm for low-emission, light-duty vehicles with new catalysts to 0.2 ppb/ppm for new diesel heavy-duty vehicles [47–50]. (Interestingly, $N_2O/CO_2$ ratios have been increasing for diesel vehicles as more advanced catalysts have been added to reduce $NO_x$ emission [48,50].) From Figure 3, we do observe some correlation between $N_2O$ and $CO_2$ but less correlation than between CO and $CO_2$. Figure 5a shows a scatter plot of excess $N_2O$ versus excess $CO_2$, defined as previously for CO, over the full time period. Again, as expected, the correlation is far less tight than for the excess CO/$CO_2$ plot. This broad distribution suggests multiple source contributions with different $N_2O$/$CO_2$ ratios, perhaps due to temporal and spatial differences in traffic distributions combined with wind patterns. Based on Figure 6a, the range of correlation is ~0.02 ppb/ppm to ~0.25 ppb/ppm, which is consistent with the range of values reported previously given above. Note that there are only a few points with the highest ratio, which arise from a very narrow plume event. The majority of points lie between 0.02 ppb/ppm and 0.09 ppb/ppm. In order to better see the different source contributions, we plot the correlation for specific plume events, identified by the green shaded region in Figure 3, in Figure 5b-c. These show much tighter correlation than for the entire time series. The slopes of the correlation plots for the two plume events are 0.021 ppb/ppm and 0.087 ppb/ppm. To better understand the potential origin of these plume events and the difference in slope between them, we perform a footprint analysis as shown in Figure 5d. A footprint shows how much a source at a given grid cell will influence the concentration observed at a given receptor location and time and thus has units of concentration/area/time. We calculate the footprints using STILT-R [51] with a spatially uniform wind field generated from the local measurement. This analysis shows that the two events likely originate from different source locations and suggests that the first event (colored blue) arises from local traffic (which consists mostly of light-duty vehicles) whereas the second event (colored green) likely originates from highways with more truck traffic, leading to the higher slope.

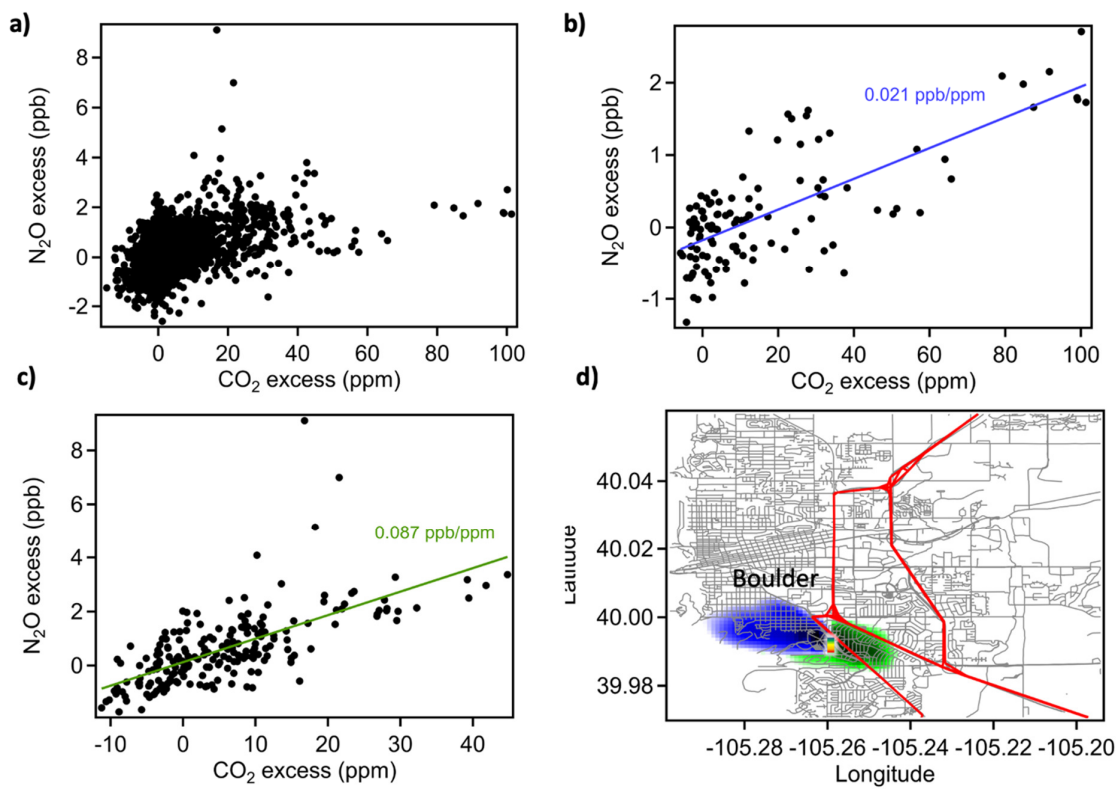

**Figure 6.** Correlation plots of excess N$_2$O versus excess CO$_2$ for (a) all data and (b-c) the green-shaded regions in Figure 3. Colored lines show the results from a line fit. (d) Map of the estimated source footprints for the plume events in (b) (blue) and (c) (green) based on the wind direction and speed. Darker colors indicate a larger footprint. Major highways are indicated in red.

## 5. Conclusions

We demonstrated open-path DCS measurements across a 600-m-long path in the 4.5 µm to 5 µm spectral region for simultaneous measurements of CO, O$_3$, N$_2$O, CO$_2$, and H$_2$O. The system operated over 5 days with an 80% uptime. The two-minute sensitivities for N$_2$O, CO, CO$_2$, O$_3$, and H$_2$O were 0.5 ppb, 0.4 ppb, 1.5 ppm, 2 ppb and 12 ppm, which are generally sufficient to capture atmospheric variations due to changes in the boundary layer or plumes from local sources. Furthermore, comparison with a co-located point sensor showed good agreement with small static offsets of 1.8 ppb, -0.88 ppm, and -2.9 ppb for CO, H$_2$O, and N$_2$O, respectively, which are attributed to uncertainties in the HITRAN data used for the retrievals.

In the future, several improvements are possible. First, the system can readily be extended to longer path lengths: we have tested the system over 2 km long paths and have demonstrated >5 km long paths in the near-infrared [7]. Second, with sensitivity improvements through higher power comb sources, lower relative-intensity noise and detector noise, and longer path lengths the system could detect additional species and isotopologues at atmospherically relevant levels. These improvements would add even more information for source apportionment and enable further applications. Similarly, with regard to air quality, future system improvements could enable high time resolution, sensitive $O_3$ measurements to develop a better understanding of the interplay between different factors influencing $O_3$ formation [52,53].

The multispecies measurement capability of the instrument has clear applications to quantifying greenhouse gas emissions and validating emission inventories, as open-path dual-comb spectroscopy can now sense all four primary greenhouse gases, $N_2O$, $CO_2$, $CH_4$ and $H_2O$. In addition, the $N_2O$ sensitivity is sufficient to enable temporally and spatially resolved $N_2O$ flux characterization using a flux gradient approach [54] from agricultural sources and wetlands, which will help to constrain $N_2O$ sources and enable monitoring of $N_2O$ emission reduction measures [36–38]. From the data here, we provide initial correlation studies of both CO and $N_2O$ with $CO_2$. Further extended deployments could provide a rich database for comparison with the NEI or other inventories and for guiding urban GHG reduction measures.

**Acknowledgements**

We acknowledge valuable comments from B.R. Washburn and A.J. Fleisher. Funding provided by NIST and the DARPA SCOUT program.